\documentclass{elsart}

\usepackage[square]{natbib}

\usepackage{graphics}
\usepackage{verbatim}
\usepackage{graphicx}
\usepackage{amssymb}
\journal{Nucl. Instr. and Meth. in Phys. Res. A}

\def\PANDA{$\overline{\mbox{\sf P}}${\sf ANDA}\ }

\begin{document}

\begin{frontmatter}

\title{Performance of HPGe Detectors in High Magnetic Fields}

\author[kph]{A. Sanchez Lorente\corauthref{cor}\thanksref{PhD}}
\ead{lorente@kph.uni-mainz.de},
\author[kph]{P. Achenbach},
\author[Torino-Poli]{M. Agnello},
\author[Torino]{T. Bressani},
\author[Torino]{S. Bufalino},
\author[KTH]{B. Cederwall},
\author[Torino]{A. Feliciello},
\author[Torino-Poli]{F. Ferro},
\author[GSI]{J. Gerl},
\author[Torino-Poli]{F. Iazzi},
\author[GSI]{M. Kavatsyuk},
\author[GSI]{I. Kojouharov},
\author[Rez]{L. Majling},
\author[Bari]{A. Pantaleo},
\author[Bari]{M. Palomba},
\author[kph]{J. Pochodzalla},
\author[Catania]{G. Raciti},
\author[GSI]{N. Saito},
\author[GSI]{T.R. Saito},
\author[GSI]{H. Schaffner},
\author[GSI]{C. Sfienti},
\author[Torino-Poli]{K. Szymanska}, \and
\author[Stockholm]{P.-E. Tegn{\'e}r}\\
\collab{(HyperGamma Collaboration)}
\address[kph]{Institut f{\"u}r Kernphysik, Johannes Gutenberg-Universit{\"a}t Mainz, Germany}
\address[Torino-Poli]{Dipartimento di Fisica del Politecnico di Torino, Italy}
\address[Torino]{Dipartimento di Fisica Sperimentale, Universit{\`a} di Torino and INFN Sezione di Torino, Italy}
\address[KTH]{Kungliga Tekniska H{\"o}gskolan (KTH), Stockholm, Sweden}
\address[GSI]{Gesellschaft f{\"u}r Schwerionenforschung (GSI), Darmstadt, Germany}
\address[Rez]{Nuclear Physics Institute, Academy of Sciences of the Czech Republic, Rez near Prague, Czech Republic}
\address[Bari]{Dipartimento di Fisica, Universit{\`a} di Bari, and INFN Sezione di Bari, Italy}
\address[Catania]{Dipartimento di Fisica, Universit{\`a} di Catania, and INFN Sezione di Catania, Italy}
\address[Stockholm]{Deparment of Physics, Stockholm University, Sweden}
\thanks[PhD]{Part of doctoral thesis.}
\corauth[cor]{Corresponding author. Tel.:
+49--6131--39--25802/25803; fax: +49--6131--39--22964.}

\begin{abstract}
A new generation of high-resolution hypernuclear
$\gamma$-spectroscopy experiments with high-purity germanium
detectors (HPGe) are presently designed at the FINUDA spectrometer
at DA$\Phi$NE, the Frascati $\phi$-factory, and at {\PANDA}, the
$\mathrm{\overline{p}p}$ hadron spectrometer at the future FAIR
facility. Both, the FINUDA and \PANDA spectrometers are built around
the target region covering a large solid angle. To maximise the
detection efficiency the HPGe detectors have to be located near the
target, and therefore they have to be operated in strong magnetic
fields ($B\approx 1\,$T). The performance of HPGe detectors in such
an environment has not been well investigated so far. In the present
work VEGA and EUROBALL Cluster HPGe detectors were tested in the
field provided by the ALADiN magnet at GSI. No significant
degradation of the energy resolution was found, and a change in the
rise time distribution of the pulses from preamplifiers was
observed. A correlation between rise time and pulse height was
observed and is used to correct the measured energy, recovering the
energy resolution almost completely. Moreover, no problems in the
electronics due to the magnetic field were observed.
\end{abstract}


\end{frontmatter}

\section{Introduction}
\label{introduction} High resolution $\gamma$-ray spectroscopy based
on high-purity germanium (HPGe) detectors represents one of the most
powerful experimental tools in nuclear physics. The introduction of
this technique led to a significant progress in the knowledge of
nuclear structure. It has been recently proven that also strangeness
nuclear physics can benefit from the same advantages: the energy
resolution of hypernuclear levels has been drastically improved from
1--2\,MeV to a few keV in FWHM~\cite{Tamura2000,Hashimoto2006}.

The success of such a technique has encouraged other groups working
on FINUDA at DA$\Phi$NE~\cite{Bressani2005} and \PANDA at
FAIR~\cite{PANDA,joput} to investigate whether this technique could
be extended and incorporated in their set-ups. The FINUDA and \PANDA
magnetic spectrometers have a cylindrical geometry and are built
around the target region covering a large solid angle
($\Omega\approx 4\pi\,$sr). To maximise the detection efficiency
HPGe arrays are to be mounted near the target region, implying the
operation of these detectors in a strong magnetic field (up to
$B\approx1\,$T). At DA$\Phi$NE, collisions between electrons and
positrons of 510 MeV lead to an abundant production of $\Phi$(1020)
mesons, decaying predominantly into low energy ($\sim16$\,MeV)
K$^+$K$^-$ pairs. The FINUDA spectrometer is centred around a set of
eight thin (0.2\,-- 0.3\,g$/$cm$^2$) nuclear targets, surrounding
the interaction point. Since the year 2003 $\Lambda$-hypernuclei
produced by $K^-$ mesons stopped in these four targets have been
studied~\cite{Agnello2005}. At \PANDA relatively low momentum
$\Xi^-$ can be produced in $\mathrm{\overline{p}p} \to \Xi^-
\overline{\Xi}^+$ or $\mathrm{\overline{p}n} \to \Xi^-
\overline{\Xi}^0$ reactions. A rather high luminosity is anticipated
to be achieved due to retaining the antiprotons in a storage ring
which will allow the use of thin targets. The associated
$\overline{\Xi}$ will undergo scattering or (in most cases)
annihilation inside the residual nucleus.  Strangeness is conserved
in the strong interaction and the annihilation products contain at
least two anti-kaons that can be used as a tag for the reaction.  In
combination with an active secondary target, high resolution
$\gamma$-ray spectroscopy of double hypernuclei and $\Omega$ atoms
will become feasible for the first time~\cite{joput,Hiyama}.

The aim of the work described in this paper is to study the
feasibility of using HPGe detectors in high magnetic fields and to
study the associated effects on their energy resolution. The pulse
shape distortion is also investigated.

\section{Germanium Detectors in Magnetic Fields}
Germanium detectors which are typically used in low energy nuclear
spectroscopy (see e.g. \cite{jubeck1,jubeck2}) are seldom operated
in magnetic fields \cite{per} and their behaviour under such
conditions is not well known. Generally the deflection of the charge
carriers in the magnetic field and the Penning effect in the vacuum
surrounding the semiconductor can play a substantial role for the
operation of large volume semiconductor devices in high magnetic
fields.

For extended detector volumes and hence long drift paths the
deflection of the charge carriers in the magnetic field may result
in a larger rise-time of the signal due to longer drift paths or
enhanced trapping and detrapping. By using standard shaping
amplifiers the output signal reflects then the interplay between
charge collection process and the transfer function of the
amplifier. In practice an enhanced charge collection time will cause
a reduction of the output signal even though the complete charge is
collected eventually. In case of trapping the timescales involved
may be significantly larger than the typical time constants of the
electronic network and this reduction is referred to as a ballistic
deficit \cite{Knoll}. Sometimes this term is also used in a more
general meaning \cite{bal1,bal2} for any decrease of the output
signal due to an enhanced signal rise-time irrespective of its
origin. In any case the associated larger fluctuations of the signal
rise-time will deteriorate the energy resolution. Trapping of charge
carriers and losses due to recombination depend on the type of the
germanium. In the case of n-type germanium detectors which are
studied in the present work trapping is expected to be less
significant than in p-type germanium detectors. Nonetheless, a
reduction of the signal may be particularly important in the case of
major radiation damages of the crystal lattice (see e.g.
\cite{rad}).

The bending force of the magnetic field causes charged particles
produced within the volume between the Ge crystal and the capsule to
spiral around the field lines. The longer travel times of the rest
gas ions may result in an enhanced Penning effect. The interaction
of electrons with the residual gas within the capsule may cause
secondary ionisation leading eventually to discharges.

The generation of an electric field perpendicular to the magnetic
field lines and the direction of the current (Hall effect) may
affect small electronic components carrying large currents. While
the wide spread use of silicon detectors and their associated
readout electronics in tracking systems demonstrates the feasibility
to operate highly integrated electronic devices in high magnetic
fields (see e.g. ~\cite{simag1,simag2,simag3,simag4}), the effect on
the electronics of high resolution devices like HPGe detectors has
not been studied yet.

\section{Experimental Details}
To verify that HPGe detectors can be safely and efficiently operated
in a high magnetic field, two different kinds of detectors have been
tested: the EUROBALL Cluster detector~\cite{Eberth1996} and the VEGA
detector~\cite{Gerl1994}.

\subsection{The EUROBALL Cluster Detector}
The EUROBALL Cluster detector consists of seven large hexagonal,
n-type, closely packed, tapered Ge crystals housed in a common
cryostat~\cite{Eberth1996}. The crystals have a length of 78\,mm and
a diameter of 70\,mm at the cylindrical back-end. To protect the
sensitive intrinsic surface of the detectors and to improve the
reliability each crystal is encapsulated. The capsules are made of
aluminium with a thickness of 0.7\,mm. The distance of the Ge
surface to the inner wall of the capsule is only 0.7\,mm which gives
a distance of $3.0-3.5$\,mm between the edges of two neighbouring
detectors in a cluster. Each capsule is hermetically sealed by
electron beam welding of the capsule lid. The vacuum is maintained
by a getter material which is active up to the temperature
of 150$^o$\,C. The cold part of the preamplifier is mounted on the
capsule lid. The detector has a typical energy resolution of
2.1\,keV (FWHM) at $1.332\,$MeV for a $^{60}$Co source using an AC
coupled preamplifier. The AC coupling was chosen in order to operate
the detector capsule on ground potential which facilitates the close
packing and the cooling of several detectors in a common cryostat.
Since the seven crystals in the cluster are identical, only three
crystals were taken as a representative for the measurements.

\subsection{The segmented Clover Detector, VEGA}
The super-segmented-clover detector VEGA~\cite{Gerl1994} consists of
four large coaxial n-type Ge crystals which are four-fold
electrically segmented. The crystals have a length of 140\,mm and a
diameter of 70\,mm. They are arranged in the configuration of a
''four-leaf'' clover, and housed in a common cryostat. The core
contact is AC coupled and the segments are DC coupled. The
preamplifier used in the VEGA detector are similar to the one
operated with the Cluster detector. Three crystals (B, C and D) and
all four segments of one of the three (crystal B) were used in the
present work. Prior to the studies in a magnetic field the energy
resolution of the detectors was measured in the laboratory with a
$^{60}$Co source to be about $2.2\,$keV (FWHM) at $1.332\,$MeV.

\begin{figure}[p]
 \centerline{\includegraphics[width=\columnwidth]{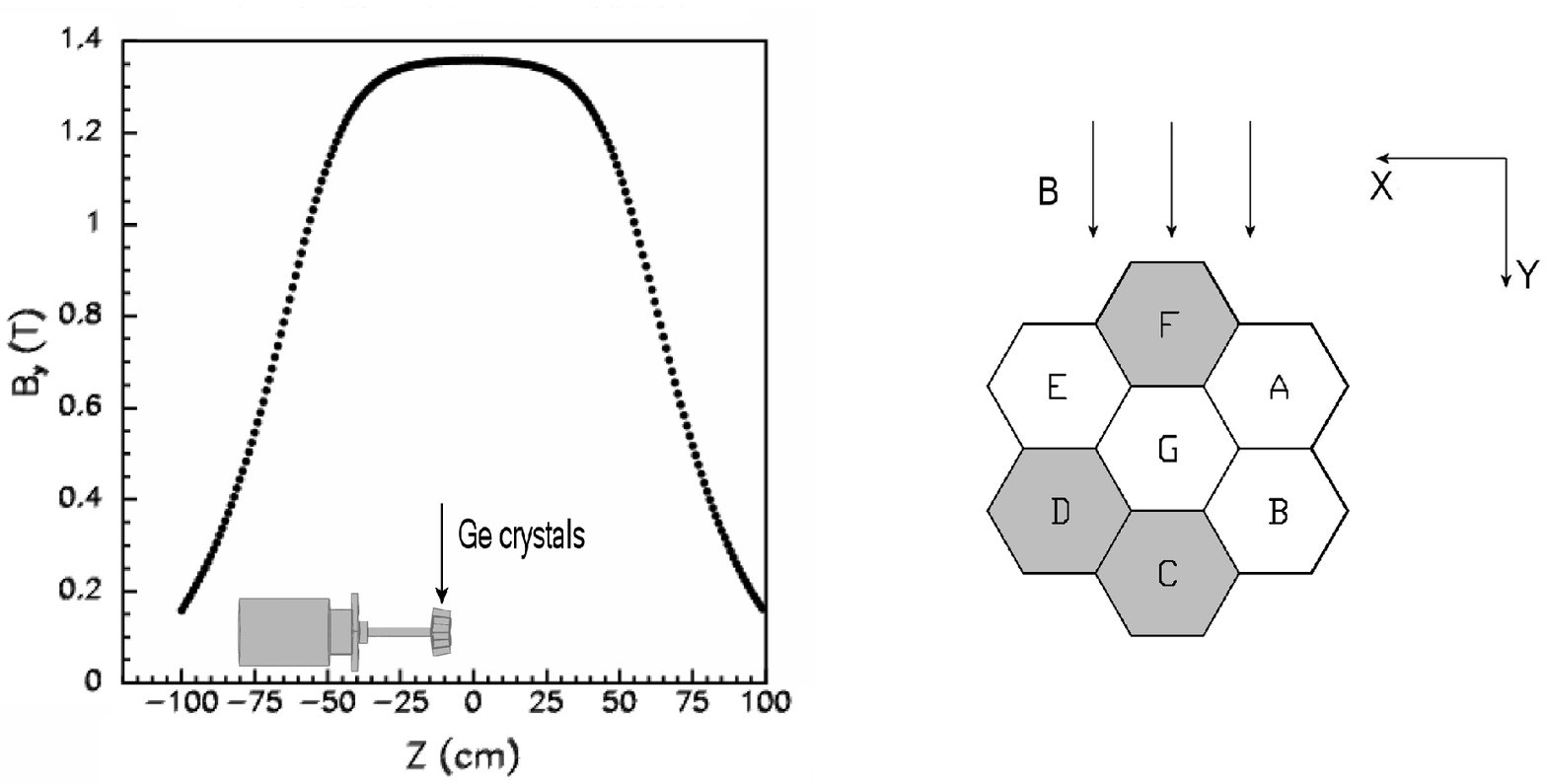}}
 \caption{Illustration of the EUROBALL Cluster orientation inside the
 ALADiN magnet.  The left panel shows the Cluster position inside the
 magnetic field and the calculated field map along the $z$-direction
 for 1800\,A coil current. The right panel shows the end-cap of the
 Cluster, where the shaded crystals F, D, and C represent those
 which have been used during the measurements.}
 \label{fig:gam:eubsetup}
\end{figure}
\begin{figure}[p]
  \centerline{\includegraphics[width=\columnwidth]{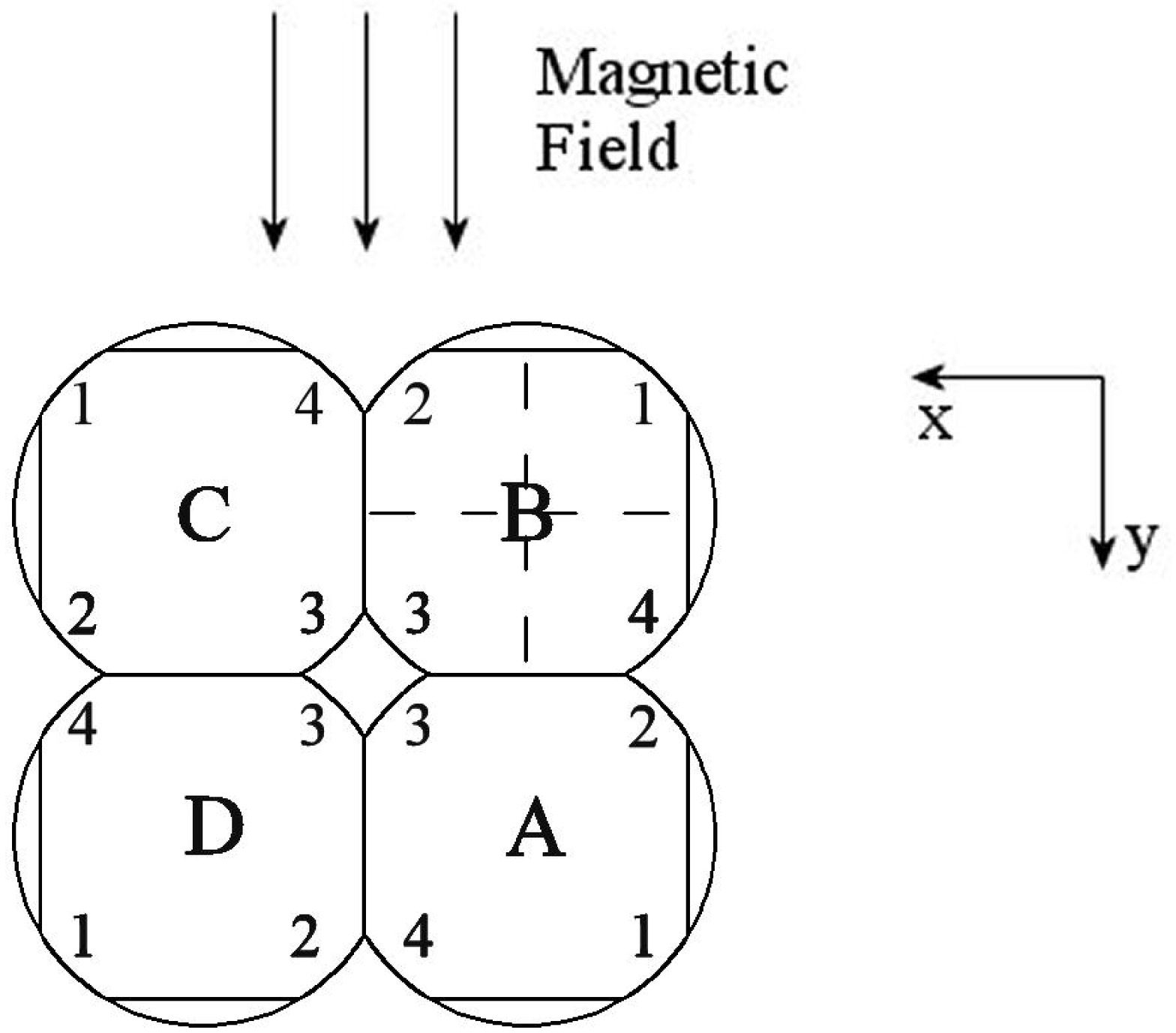}}
  \caption{Illustration of the end-cap of the VEGA detector in clover
  geometry and its orientation inside of the ALADiN magnet. All four
  segments labeled by the numbers 1, 2, 3 and 4 of channel~B were read out
  individually with a Flash-ADC.}
  \label{fig:gam:vegB}
\end{figure}

\subsection{Experimental Set-up}
Two series of measurements were performed using the ALADiN dipole
magnet~\cite{ALADiN1988}. For both series the HPGe detectors and a
$^{60}$Co $\gamma$-ray source, with an activity of 370\,kBq, were
positioned inside the magnet with the source placed in front of each
detector. The ALADiN magnet aperture of $1.5 \times 0.5$\,m$^2$
restricts to place a detector with its geometrical axis in the
horizontal plane of the magnet. The direction of the magnetic field
lines was perpendicular to the geometrical detector axis as shown in
Fig.~\ref{fig:gam:eubsetup}\,(right). The magnetic field is maximal
in the centre of the magnet and decreases along the $z$-direction as
illustrated in Fig.~\ref{fig:gam:eubsetup}\,(left). The detector
end-caps have been placed as close as possible to the centre of the
magnet in order to expose the germanium crystals to the highest
magnetic field (about 7\,cm from the centre as shown in
Fig.~\ref{fig:gam:eubsetup}).

The $^{60}$Co source was placed at a distance of 27\,cm and 20\,cm
away from the end-cap of the EUROBALL Cluster and VEGA detector,
respectively. Fig.~\ref{fig:gam:eubsetup} shows the scheme of the
crystals inside the Cluster detector in the right panel. For the
measurements with the EUROBALL Cluster three channels (C, D, F) out
of the seven Ge crystals of the Cluster were used. Channels~C and~D
were fed with a voltage of 4000\,V and channel F with 3500\,V. For
the measurements with the VEGA detector data from three (B, C, D) of
the four Ge crystals were analysed. The geometry of the crystals of
VEGA detector inside the magnet is shown in Fig.~\ref{fig:gam:vegB}.
They have been fed with voltages of 4000\,V. The measurements can be
divided in two groups: measurements done without magnetic field and
those in which the magnetic field was tuned to 0.3\,T, 0.6\,T,\,0.9
T, and 1.4\,T for the EUROBALL Cluster detector and to 0.6\,T,
1.1\,T, 1.4\,T, and 1.6\,T for the VEGA detector. For each EUROBALL
Cluster channel (C, D, and F) one of the preamplifier outputs was
split in two signals. One signal was fed to a spectroscopy amplifier
(Ortec 572) with 3\,$\mu$s shaping time, which output was digitised
by an Analog-to-Digital converter (ADC Silena 4418/V, 8 channel, 12
bit resolution) and the second signal was fed to a VME 100\,MHz
Flash-ADC (FADC SIS3300, 8 channel, 12 bit resolution). The other
output from the preamplifier was sent to a Timing-Filter-Amplifier
(TFA Ortec 474), whose output was discriminated by a
Constant-Fraction-Discriminator (Ortec CF 8000) to be used as a
timing signal. The trigger was formed by a logic OR of the outputs
from the three CFD channels. The ADC was read out via CAMAC bus.

Since the VEGA crystals are electrically segmented, the readout
electronics differs from that corresponding to an EUROBALL Cluster
in the trigger signal. One output of the preamplifier (core signal)
for each channel was split in two branches as it was done for the
EUROBALL Cluster set-up.  Those preamplifier outputs corresponding
to the the four segments of channel~B were fed to the FADC. The
trigger is formed from a coincidence of the logic OR of the CFD
outputs of the four segments and an external trigger determined from
the central core signal of channel~B.

\begin{figure}[p]
  \centerline{\includegraphics[width=\columnwidth]{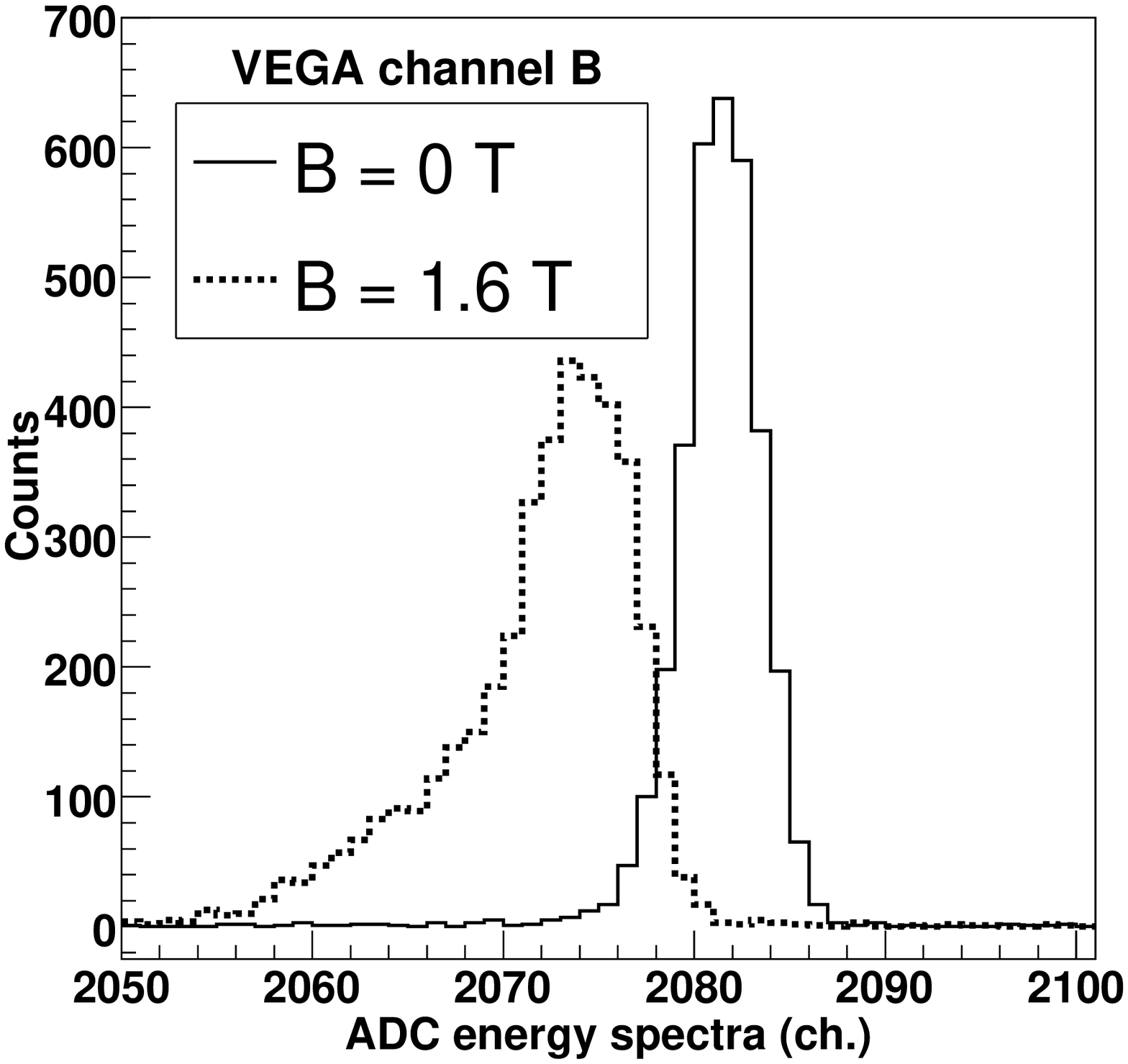}}
  \caption{Pulse height spectra for the 1.332\,MeV $^{60}$Co
  $\gamma$-ray line measured with VEGA channel~B at
  different values of the magnetic field. The continuous line presents
  the pulse height spectrum for the measurement without magnetic field and the
  dashed line the pulse height spectrum at a field of 1.6\,T.}
  \label{ADC_spectra}
\end{figure}
\begin{figure}[p]
  \centerline{\includegraphics[width=1.2\columnwidth]{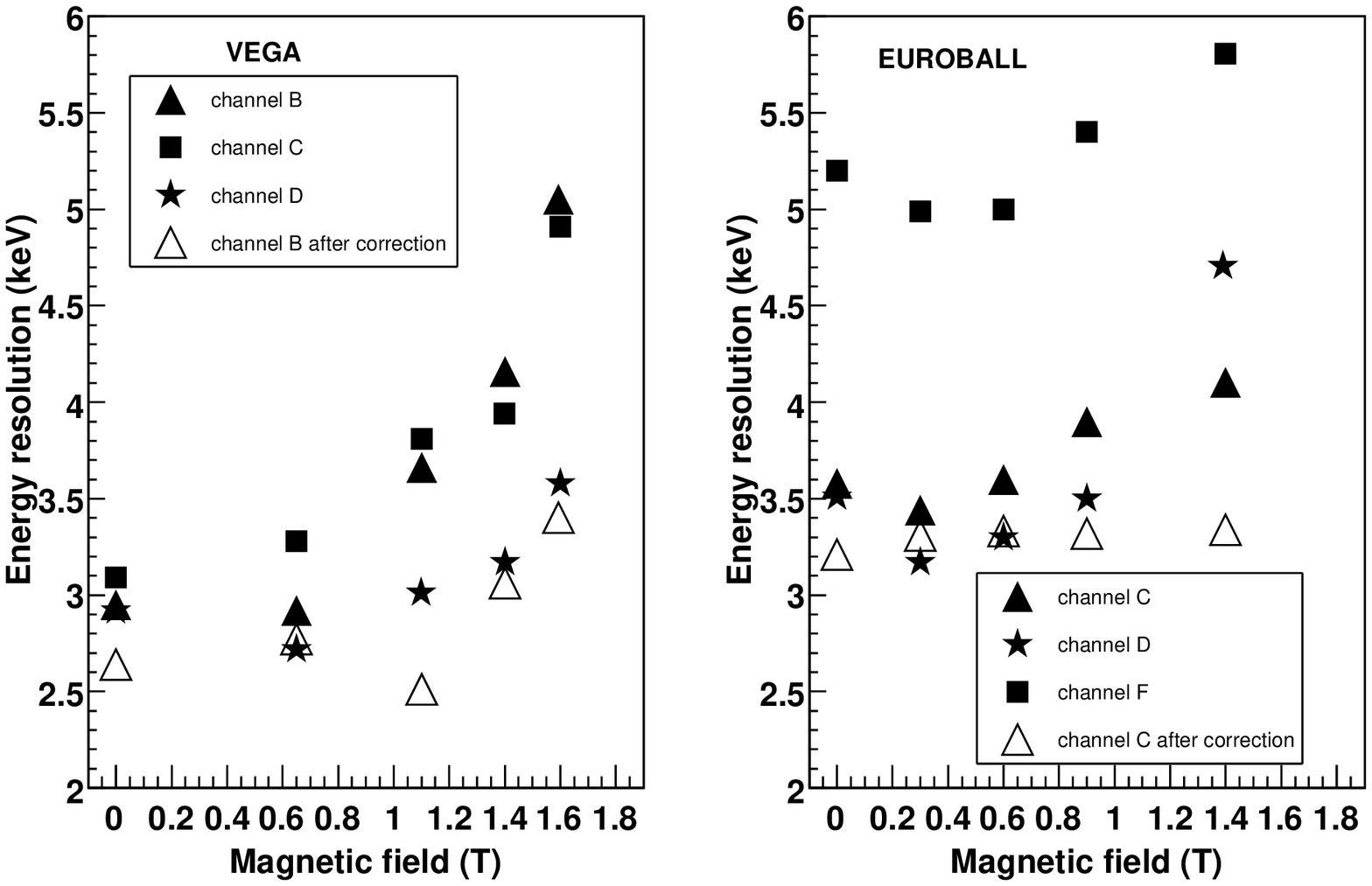}}
  \caption{Measured energy resolution (FWHM) for the 1.332\,MeV
  $\gamma$-line of VEGA and EUROBALL Cluster detectors in a magnetic
  field $B$. The resolution of EUROBALL channel~F was worse because of noise
  pick-up. The empty symbols represent the improved energy resolution
  after the correction of the energy spectra for channel B and C, respectively(see text).
  Statistical errors for the values are smaller than the symbol size.}
  \label{energyres}
\end{figure}

\section{Data Analysis and Results}
The analysis presented here is focused on the determination of the
energy resolution from the pulse height spectra of both detectors by
using conventional analogue electronics and on the study of the
dependence of the pulse shape sampled by an FADC in magnetic fields.
A detailed study on HPGe detectors operating with high rates in
magnetic fields, based on the observation of pulse shapes, will be
published in a forthcoming paper.

The energy resolution was extracted from the pulse height spectra by
parameterising the 1.332\,MeV $\gamma$-ray full-absorption peak from
a $^{60}$Co source. From the parameterised line shape the value of
Full-Width Half-Maximum (FWHM) was extracted. Two different methods
were used to extract the energy resolution of the detectors
depending on whether the spectra were measured in a magnetic field
or not. Since in absence of a magnetic field the line shape of all
detectors is very close to a single Gaussian distribution, the pulse
height spectra have been parametrised by Gaussian function. In case
the magnetic field was non-zero, the convolution of a Gaussian
distribution and an exponential decay function was chosen to
describe also the tail on the low-energy side of the peak. The
observed line shape was fitted by a full absorption-peak
superimposed on a quadratic background.

\begin{figure}[p]
  \centerline{\includegraphics[width=\columnwidth]{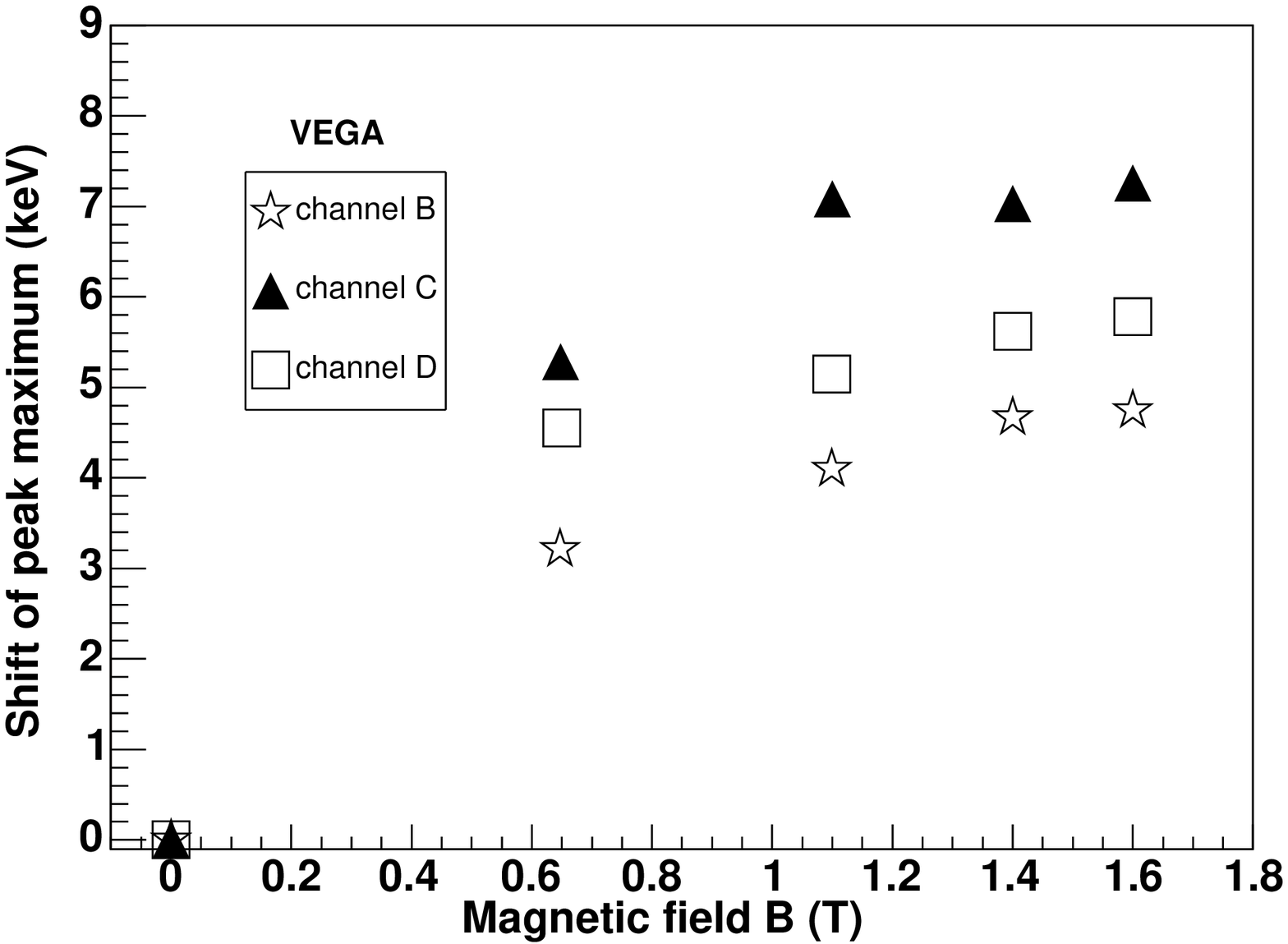}}
  \caption{Shift to smaller energies of the 1.332\,MeV
  $\gamma$-line peak maximum of the VEGA detector in a magnetic
  field $B$.}
  \label{peakshifts}
\end{figure}
\begin{figure}[p]
  \centerline{\includegraphics[width=\columnwidth]{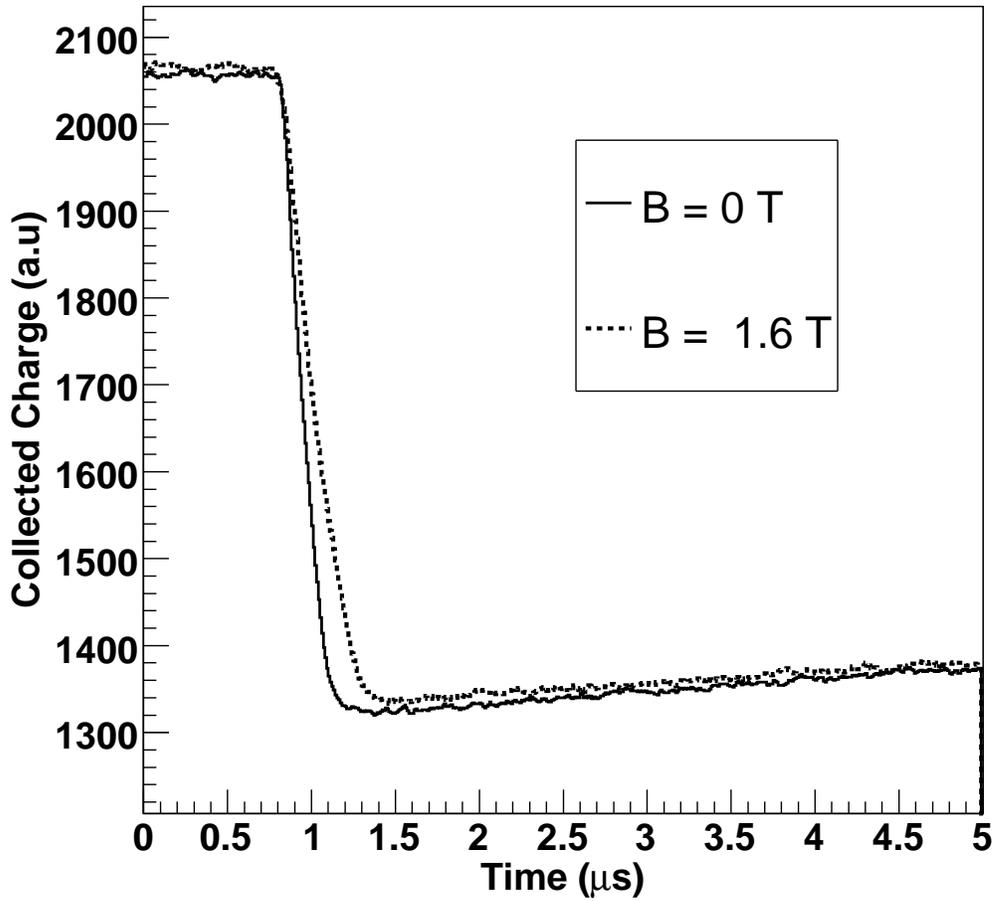}}
  \caption{Averaged preamplifier outputs signals of VEGA channel~B for the
1.332\,MeV $\gamma$-ray measured at a magnetic field of $B= 0$\,T (solid line)
and at $B = 1.6$\,T (dashed line). The two pulse shapes signals have been obtained
by averaging over 5000 events each.}
  \label{pulseshape}
\end{figure}
\begin{figure}[p]
  \centerline{\includegraphics[width=0.8\columnwidth]{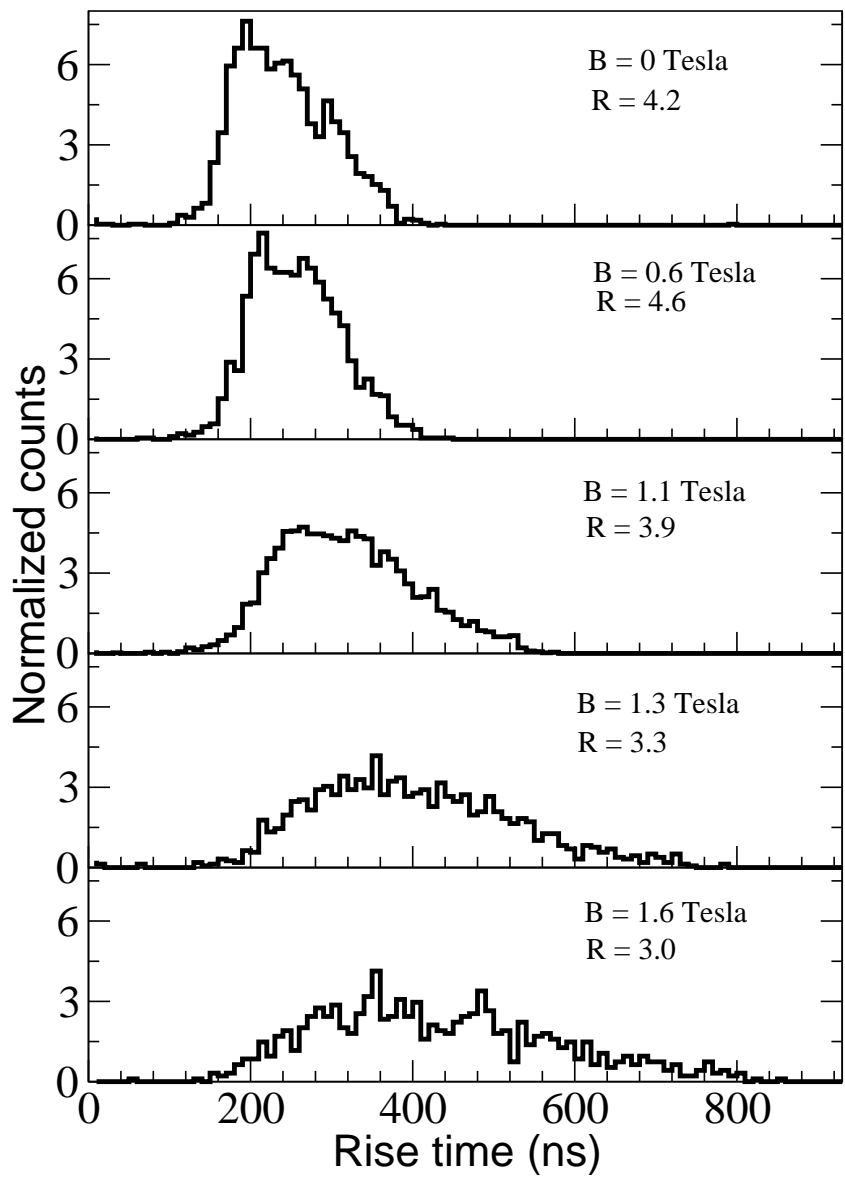}}
  \caption{Measured rise time distribution for the 1.332\,MeV
  $\gamma$-line of one of the segments of VEGA channel~B in a magnetic
  field $B$. $R$ gives the ratio between the mean value and the RMS value
  of the rise time distributions. Each distribution has been normalized by its integral.}
  \label{risetimes}
\end{figure}
\begin{figure}[p]
  \centerline{\includegraphics[width=1.2\columnwidth]{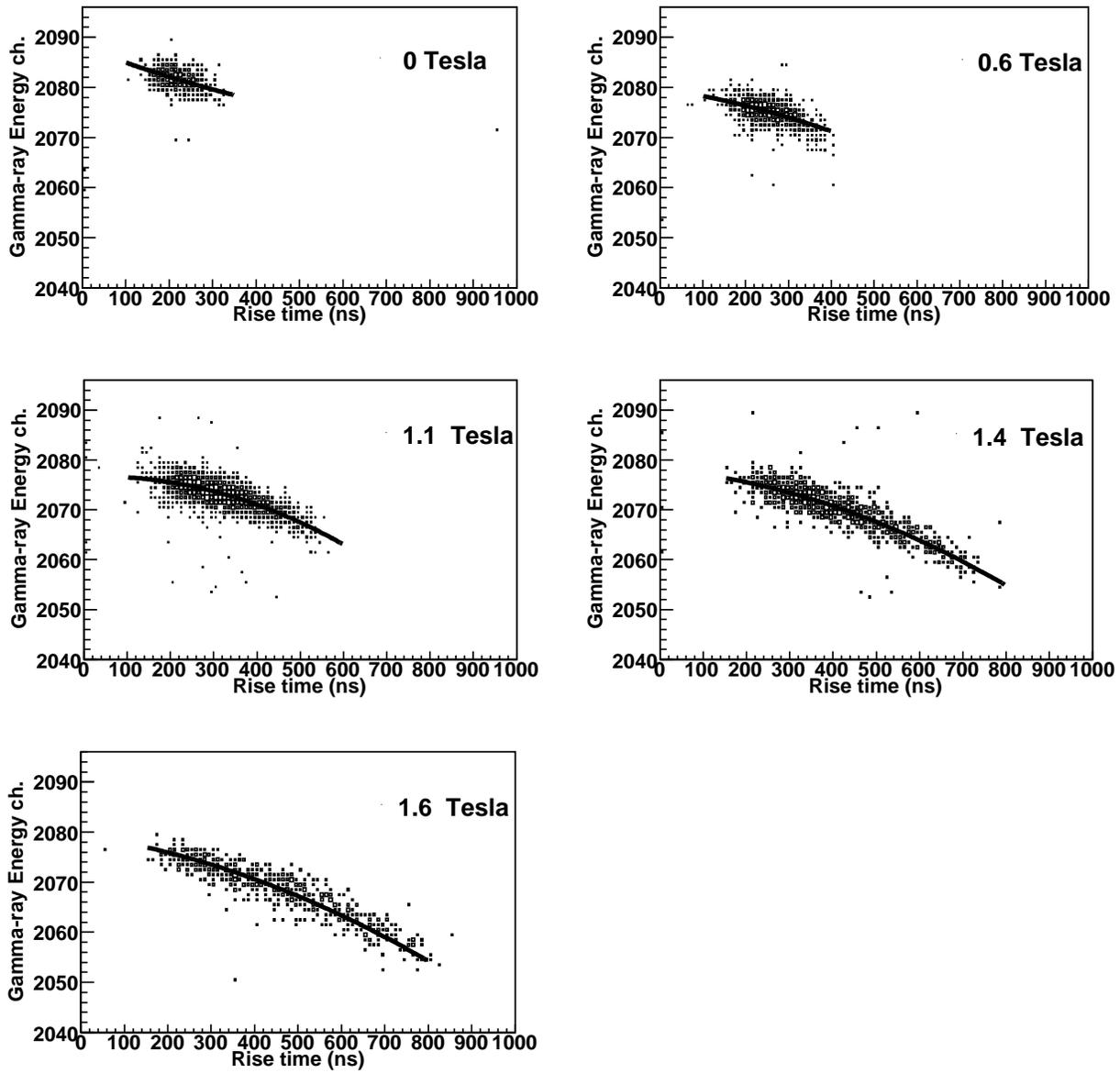}}
  \caption{Correlation between the energy in channels and the rise time for the 1.332\,MeV
  $\gamma$-rays of one of the segments of VEGA channel~B in a magnetic
  field. The line represents the parabolic fit function of each
  distribution at different values of the magnetic field.
  }
  \label{T90Adc}
\end{figure}
\begin{figure}[p]
  \centerline{\includegraphics[width=\columnwidth]{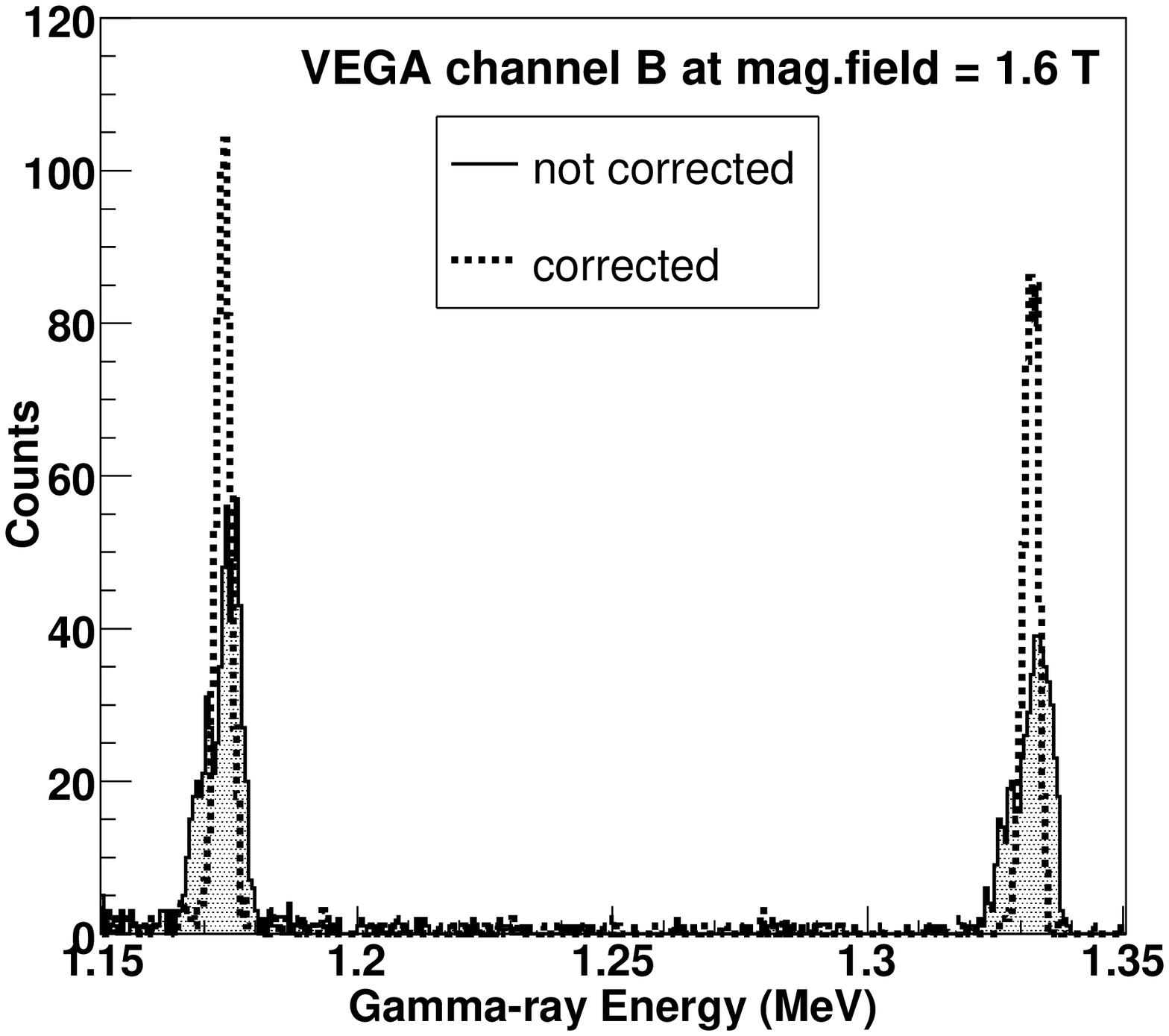}}
  \caption{Two $\gamma$-ray energy spectra for $^{60}$Co measured with VEGA channel~B at
  maximum value of the magnetic field. The hashed spectrum presents
  the pulse height spectrum for the measurement without correction and the
  dashed line the corrected pulse height spectrum at a field of 1.6\,T.
  Both spectra have been individually calibrated.}
  \label{calADC_spectra}
\end{figure}

Pulse height spectra taken with a shaping amplifier are presented in
Fig.~\ref{ADC_spectra} for the 1.332\,MeV $^{60}$Co $\gamma$-ray
line. Obviously, the line shape at $B = 1.6$\,T  (dotted histogram)
is significantly different from the line shape measured without
field (solid histogram). The influence of the magnetic field causes
a tail on the low-energy side. The dependence of the energy
resolution (FWHM) on the strength of the magnetic field is shown for
the EUROBALL Cluster as well as for the VEGA detector in
Fig.~\ref{energyres}. The energy resolution of one of the EUROBALL
Cluster crystals is worse than the resolution of the two other
crystals of the same detector because of pick-up noise in its
electronic readout. In addition, the peak maximum of the
$\gamma$-ray line shifts towards smaller energies with increasing
field strength as shown in Fig.~\ref{peakshifts}.

In order to clarify the question whether the shifts of the energy
peak are due to a loss of charge carriers or whether this shift is
caused by the interplay between an increased charge collection time
in the Ge detector and the transfer function we have studied the
FADC signals. Fig.~\ref{pulseshape} shows the averaged pulse shape
signals for 1.332\,MeV $\gamma$-rays measured at zero magnetic field
(solid line) and at $B = 1.6$\,T (dashed line). The two signals have
been obtained by averaging over 5000 events each. Within the two
time intervals of [0, 0.5\,$\mu$s] and [3\,$\mu$s, 4\,$\mu$s] one
finds mean amplitudes of 2056.7 (2065.7) and 1353.9 (1362.9) for $B
= 0$\,T ($B = 1.6$\,T), respectively. Despite a very different shape
of the rising edge, the difference between the initial baseline and
the signal level at later times stays nearly constant at a value of
about 702.8 . This indicates that the recombination or long-time
trapping of charges cannot account for the reduced energy signals
presented in Fig.~\ref{peakshifts}.

As can be seen already in Fig.~\ref{pulseshape}, the pulse shape
is modified by the magnetic field. Fig.~\ref{risetimes}
shows the distribution of the rise time (defined as the time it
takes for the pulse to rise from 10\,\% to 90\,\% of its full
amplitude) for different values of the magnetic field for the VEGA
detector. A significant change of its mean value by approximately
$200$\,ns and a broadening of the distribution can be observed. The
ratio $R$ between the mean rise time and the root mean square value
decreases from values above $4$ at low magnetic fields to $3.0$ at $B
= 1.6$\,T. This behavior is opposite to what is expected for a
purely diffusive motion of the charge carriers. In the latter case
an increase of R proportional to the square root of the rise time
would have been expected.

The simultaneous large shift and the broadening of the rise time
distributions on one hand, and the rather similar asymptotic values
of the pulses seen in Fig.~\ref{pulseshape} on the other hand
suggest that the incomplete signal integration by the main amplifier
is the main origin of the observed energy shift. To verify the
latter conjecture, the dependence of the amplifier transfer function
on the variation of the rise time has been investigated with a pulse
generator. This study confirmed that the rise-time variations are
the main source of the energy shift seen in Fig.~\ref{peakshifts}.

In order to explore the possibility to correct the shift of the
pulse height by measuring the rise time event-by-event we show in
Fig.~\ref{T90Adc} the correlation of the deduced $\gamma$-ray energy
and the rise time for different magnetic fields. These distributions
have been obtained event-by-event for the 1.332\,MeV $\gamma$-line
of one of the segments of VEGA channel~B. For all magnetic fields
the low-energy tail of the $\gamma$-ray peak (see Fig.
\ref{ADC_spectra}) is associated with an increased rise time. The
strong correlations observed in Fig.~\ref{T90Adc} provides a
characterisation of the amplifier at different values of the rise
time. The fit of a parabolic function to the correlation (see dark
lines in Fig.~\ref{T90Adc}) enables the correction of the energy
spectra event-by-event for different magnetic fields. The correction
functions are similar for all measurements at non-zero magnetic
field, but they are shifted relative to that of the measurement at
$B = 0$\,T. Presently the origin of this strong shift is not
understood.

Fig.~\ref{calADC_spectra} shows the corrected $\gamma$-ray energy
spectra (dashed line) for VEGA channel~B at 1.6\,T in comparison to
the one without correction (hashed). A significant improvement in
the peak shape has been obtained. After applying this correction of
the pulse height spectra, an improvement on the energy resolution
for VEGA channel~B as well as for EUROBALL Cluster channel~C has
been achieved as shown in Fig.~\ref{energyres}. The open triangles
in Fig.~\ref{energyres} represent the energy resolution for both
detectors after this correction.

\section{Discussion and Conclusion}
Two important effects have been observed by operating HPGe detectors
in high magnetic fields: a small degradation of the energy
resolution and a change of the pulse shape of the preamplifier
signal.

All crystals of both detectors show a similar behaviour in the
magnetic field with an energy resolution degradation of about
1--2\,keV. However, the resolution is still sufficient to perform
$\gamma$-ray spectroscopy on hypernuclei. The asymmetry of the line
shape appears as a low-energy tail in the pulse height spectra in
Fig.~\ref{ADC_spectra}. Moreover, the mean value of the energy
spectrum exhibits a shift to low energies as shown in
Fig.~\ref{peakshifts}. The measurements have been performed over a
period of two days without observing any problems in the electronics
or sparking effects. After the measurements the original energy
resolution was recovered.

A significant shift and broadening of the rise times distribution
has been observed in the presence of a high magnetic field. This
observation reflects the effect of the magnetic field on the charge
collection process itself.

The observation of a strong correlation between the measured pulse
height measured with analogue electronics and the rise time measured
by a FADC for various magnetic field strengths reveals the change in
rise time as the major contribution to the degradation of the energy
resolution. Employing this correlation for a rise time correction of
the energy allows to almost recover the original energy resolution.
The remaining degradation at fields larger than 1\,T amounts to
approximately 0.5\,keV.

One could expect a dependence on the orientation of the detector
with respect to the magnetic field. In our case a complete test of
the orientation of the detector could not be carried out because of
technical limitations~\cite{Sanchez2004}: the aperture of the ALADiN
magnet does not allow to freely rotate the axis of the detector.
Nonetheless, the geometry of the setup used in the present study
reflects the operating conditions of HPGe detectors in the future
FINUDA experiment, since the detectors will be set up almost
perpendicular to the direction of the magnetic field. On the other
hand, the set-up of the HPGe detectors in the future \PANDA
experiment requires a further test with the magnetic field
orientation almost parallel the detector axis.

The HPGe detectors were found to operate well in magnetic field
conditions similar to those to be expected in future hypernuclear
experiments at FINUDA and at \PANDA. However, the detectors used in
the present work are coupled to huge dewars for their cooling with
liquid nitrogen. These dewars have been the main obstacle for the
study of the dependence of the energy resolution on the orientation
of the detector in the magnetic field of ALADiN. Furthermore they
affect the detector integration in both the FINUDA and \PANDA
magnetic spectrometers. In order to circumvent these problems an
electromechanical cooling system coupled to few encapsulated HPGe
crystals is currently under development.

\section*{Acknowledgements}
This research is part of the EU integrated infrastructure initiative
Hadron Physics I3HP under contract number RII3-CT-2004-506078. We
acknowledge financial support from the Bundesministerium f{\"u}r
Bildung und Forschung (bmb+f) under contract number 06MZ176. T. R.
Saito acknowledges the grant from the Helmholtz association as
Helmholtz-University Young Investigators Group VH-NG-239. The
authors are very grateful to the EUROBALL Owners Committee and the
RISING collaboration for the availability of the Cluster capsules to
study HPGe crystal behaviour in magnetic fields.

\end{document}